\documentclass[pop,fleqn]{w-art}
\usepackage{times}
\usepackage{w-thm}
\usepackage{graphicx}

\providecommand{\href}[2]{#2}
\begin{document}

\DOIsuffix{theDOIsuffix}


\pagespan{1}{}


\title{Fluxes, moduli fixing and MSSM-like vacua in Type IIA String Theory}

\author[P. G. C\'amara ]{Pablo G. C\'amara\inst{1,}
  \footnote{E-mail:~\textsf{pablo.gonzalez@uam.es}}}
\address[\inst{1}]{Departamento de F\'{\i}sica Te\'orica C-XI and 
Instituto de F\'{\i}sica Te\'orica C-XVI,\\ Universidad Aut\'onoma 
de Madrid, Cantoblanco, 28049 Madrid, Spain}
\begin{abstract}
We review some of the features of Type IIA compactifications in the
presence of fluxes. In particular, the case of
$T^6/(\Omega (-1)^{F_L} \sigma)$ orientifolds with RR, NS and metric
fluxes is considered. This has revealed to possess remarkable properties such
as vacua with all the closed string moduli stabilized, null or
negative contributions to the RR tadpoles or supersymmetry on the
branes enforced by the closed string background. In this way,
Type IIA compactifications with non trivial fluxes seem to constitute 
a new window into the building of semi-realistic models in String Theory.
\end{abstract}
\hspace{\stretch{1}}IFT-UAM-CSIC-05-54\\

\maketitle

\section{Introduction}

One of the most pressing problems in string theory is the issue of
moduli fixing. There is great arbitrariness in the
compactification of the theory to four dimensions, usually
parametrized in terms of a broad moduli space. For trivial 
supergravity backgrounds, the scalar potential in the moduli space 
remains flat and the low energy spectrum has plenty of massless 
scalars. However, if Superstring Theory pretends to be a unified 
theory of all particles and interactions, it should provide us with 
a mechanism which lifts these scalars from the low energy spectrum.

Recently there has been remarkable progress by considering the 
possibility of having diluted backgrounds of the RR and NSNS forms 
\cite{grana}. From the effective supergravity perspective, these 
induce non-trivial superpotentials which stabilize some (or even 
all) of the original moduli so they become massive. Most of the work 
up to now has been done in the context of Type IIB String Theory with 
three form fluxes. In that case, the minima of the superpotentials 
are associated to warped solutions of Type IIB supergravity \cite{gkp} 
and the backreaction of the fluxes represents a conformal deformation 
of the Calabi Yau metric. Thus, one can fairly consider the moduli 
space in the presence of diluted fluxes as a smooth deformation of the 
original one.

Unfortunately, all of these models correspond to no-scale solutions on 
which the K\"ahler moduli have to be stabilized through some alternative 
mechanism, such as non-perturbative effects which are always difficult 
to deal with. Moreover, at the minimum of the scalar potential, the fluxes
contribute to the RR tadpoles with the same sign as D3-branes do and one 
usually is enforced to take manifolds with large Euler numbers in order 
to fulfill the tadpole conditions and at the same time stabilize the 
moduli at large values where the $\alpha'$ and $g_s$ corrections are under 
control.

In the last couple of years analogous setups have been considered
in the context of Type IIA String Theory \cite{kashani}-\cite{kounnas}. 
In that case, one can switch on fluxes in both the even and the odd 
dimensional cycles of the internal manifold so the induced superpotential 
in principle depends on both the complex structure and the K\"ahler moduli. 

Although the ten dimensional supergravity description of this kind of vacua 
is not always well understood, in the limit of diluted fluxes the scale 
of the masses induced by the fluxes is much smaller than the Kaluza-Klein scale
and thus to consider the moduli space of the fluxed manifold as a smooth 
deformation of the original space still seems to be a fairly good approximation.

The particular case of $T^6/(\Omega (-1)^{F_L} \sigma)$ orientifolds has been 
recently analyzed in \cite{cfi}. In that case, the orientifold projection 
allows in addition for some set of metric fluxes corresponding to the gauging 
of some of the original isometries of the torus \cite{kounnas}-\cite{hull}. This 
renders the manifold into a twisted torus with reduced homology. This kind of 
geometric deformations very often survive to the action of an orbifold 
group, even though this kills all the original isometries, so one expects to 
have similar deformations in more generic half-flat manifolds.

Type IIA compactifications with fluxes and torsion have revealed to possess 
several interesting properties for model building. First of all, the fluxes 
can contribute to the tadpoles with any sign or even not contribute. This 
gives additional freedom for cancelling the RR tadpoles in semirealistic models. 
In addition, one can find AdS vacua with all the closed string moduli 
stabilized and on which the supersymmetry on the branes is enforced by the 
background fluxes.

Here we will briefly overview some of the features found in \cite{cfi} for Type 
IIA $T^6/(\Omega (-1)^{F_L} \sigma)$ orientifolds in presence of fluxes. Thus, 
in section 2 we will review the moduli space of Type IIA toroidal orientifolds 
and the possibility of having metric fluxes. In section 3 the vacuum structure 
of the scalar potentials induced by the fluxes will be discussed. Section 4 will 
be devoted to the consistency conditions of stacks of intersecting D6-branes in 
these vacua. Finally, a concrete example of $N=1$ MSSM-like vacua with all the moduli
stabilized in AdS will be presented in section 5 and some last comments will be 
made in section 6.

\section{Type IIA orientifolds and twisted tori}

We will concentrate in the case of Type IIA
compactified to four dimensions in a simple $T^6/(\Omega
(-1)^{F_L} \sigma)$ orientifold, where $\Omega_P$ is the
world-sheet parity operator, $(-1)^{F_L}$ is the space-time
fermionic number for the left-movers and $\sigma$ is an order two
involution of the 6-torus acting in the K\"ahler 2-form and in the
holomorphic 3-form respectively as
\begin{equation*}
\sigma(J)=-J \quad ; \quad \sigma(\Omega)=\Omega^*
\end{equation*}

Moreover, we will assume the 6-torus to be factorized. Then, a
suitable cohomology basis will be given by a set of 3-forms
$\{\alpha_i,\beta_i\}$ with $i=0\ldots 3$, a set of
$\sigma$-odd (1,1)-forms $\{\omega_a\}$ and its Poincar\'e dual set of
$\sigma$-even (2,2)-forms $\{\tilde{\omega}_a\}$, with $a=1\ldots 3$. The
3-forms $\alpha_i$ ($\beta_i$) are taken in such a way that are even (odd) 
under the orientifold involution and $\int_{T^6}\alpha_i\wedge \beta_j=\delta_{ij}$. 
The fixed point of $\sigma$ corresponds to $M_4\times [\alpha_0]$  and it 
is the locus of 16 O6-planes.

It has been shown \cite{louis} that the moduli space for this kind of
orientifold compactifications is better described in terms of
the complexified forms
\begin{equation}
J_c=B+iJ \quad ; \quad \Omega_c=C_3+i\textrm{Re}(C\Omega)
\end{equation}
with $B$ the NSNS 2-form, $C_3$ the RR 3-form and $C$ a compensator field 
specified by
\begin{equation}
C=\sqrt{\textrm{Vol($T^6$)}}e^{-\phi}e^{K_{cs}/2} \quad ; \quad K_{cs}=
-\textrm{log}\left[-\frac{i}{8}\int_{T^6}\Omega\wedge\Omega^*\right]
\end{equation}

The moduli parameters then can be obtained by expanding the above
forms in the corresponding cohomology basis
\begin{equation}
U_l=i\int_{T^6}\Omega_c\wedge \beta_l \quad ; \quad T_a=
-i\int_{T^6}J_c\wedge\tilde{\omega}_a
\end{equation}
Here $T_a$ parametrize the possible K\"ahler deformations of the 
torus whereas $U_l$ with $l\neq 0$ correspond to complex structure 
deformations. $S\equiv -U_0$ is the usual moduli for the dilaton.

One can define then a metric in the moduli space through the K\"ahler 
potential
\begin{equation}
K=-\textrm{log}\left[\frac{4}{3}\int J\wedge J\wedge J\right]-
\textrm{log}~\sqrt{\textrm{Vol($T^6$)}}e^{-\phi}
\end{equation}
which in terms of the above moduli parameters takes the standard 
logarithmic form for toroidal compactifications
\begin{equation}
K=-\textrm{log}(S+S^*)-\sum_{i=1}^3\textrm{log}(T_i+T_i^*)-
\sum_{i=1}^3\textrm{log}(U_i+U_i^*)\label{kahler}
\end{equation}

As advanced in the introduction, for Type IIA
compactifications the orientifold projection may allow in addition for
some geometric deformations of the original manifold which render it
into a half-flat manifold. In particular,
one can introduce a set of metric fluxes
$\omega^k_{pq}$ in the torus
\begin{equation}
ds^2=\sum_k(dx^k+\omega^k_{pq}x^qdx^p)^2
\end{equation}
so the tangent 1-forms $\eta^i$ are no longer closed forms
\begin{equation}
d\eta^k=-\frac{1}{2}\omega^k_{pq}\eta^p\wedge \eta^q \label{metri}
\end{equation}
Due to this, some of the original cycles of the torus will disappear from 
the homology in the presence of metric fluxes.

We will consider here the most general set of metric fluxes compatible with 
the symmetry of a factorable $T^6$ and the orientifold projection. This is 
given by
\begin{equation}
\begin{pmatrix}a_1\\ a_2\\ a_3\end{pmatrix}=
\begin{pmatrix}\omega^1_{56}\\ 
\omega^2_{64}\\ \omega^3_{45}\end{pmatrix}\quad ; \quad 
\begin{pmatrix}b_{11}&b_{12}&b_{13}\\ b_{21}&b_{22}&b_{23}\\ b_{31}&b_{32}&b_{33}\end{pmatrix}
=\begin{pmatrix}-\omega^1_{23}&\omega^4_{53}&\omega^4_{26}\\
\omega^5_{34}&-\omega^2_{31}&\omega^5_{61}\\
\omega^6_{42}&\omega^6_{15}&-\omega^3_{12}\end{pmatrix}
\end{equation}

It has been shown (see e.g. \cite{kalop}) that toroidal compactifications 
with metric fluxes are equivalent to compactification on a twisted torus, 
on which some of the isometries $Z_i$ of the original torus have been 
gauged. Then, for consistency of the twisted torus structure \cite{pope}, 
the metric fluxes in general will be quantized and the $a_i$ and $b_{ij}$ 
parameters will only take integer values. This could be in part expected, 
since some of the metric fluxes are related to Type IIB NSNS fluxes by 
T-duality \cite{micu,new}.

On the other hand, either from the Jacobi identity of the algebra engendered 
by the isometries or from the Bianchi identity of eq.(\ref{metri}), one
finds that the metric fluxes must satisfy
\begin{equation}
\omega^p_{[mn}\omega^k_{r]p}=0 \label{jacobi}
\end{equation}

This constrains the set of metric fluxes which one can switch on\footnote{It can 
further be shown that $\omega^p_{pn}=0$ \cite{scherk}. However this is 
automatically satisfied by the set of metric fluxes allowed by the orientifold 
projection.}. Concretely, one has
\begin{align}
b_{ij}a_j+b_{jj}a_i&=0 \quad i\neq j\nonumber\\
b_{ik}b_{kj}+b_{kk}b_{ij}&=0 \quad i\neq j \neq k\label{constra}
\end{align}

We would like now to consider the addition of non-trivial background fluxes 
for the Type IIA NSNS and RR forms. This will be done in the following section.

\section{Vacuum structure}

It has been shown by applying gauged supergravity techniques \cite{kounnas}-\cite{hull}
or directly by dimensional reduction of the ten dimensional supergravity 
action \cite{kashani}-\cite{dwolf} that, from the point of view of the four 
dimensional $N=1$ effective theory, a background for the NSNS and the RR Type IIA 
field strengths induces a non trivial superpotential of the form
\begin{equation}
W=\int_{T^6}[\Omega_c\wedge (\overline{H}_3+dJ_c)+
e^{J_c}\wedge \overline{F}_{RR}] \label{super}
\end{equation}
where $\overline{F}_{RR}$ represents a formal sum over the RR
fluxes.

We will consider therefore a general background of the NSNS 3-form and the 
even rank RR forms and we will expand it in the cohomology basis of 
the factorized torus
\begin{align}
\overline{H}_3&=\sum_{l=0}^3h_l\beta_l &
\overline{F}_0&=-m; & \overline{F}_6&=e_0\alpha_0\wedge\beta_0&
\overline{F}_2&=\sum_{a=1}^3q_a\omega_a; & \overline{F}_4&=
\sum_{a=1}^3e_a\tilde{\omega}_a
\end{align}

Since the fluxes are quantized over the corresponding $p$-cycles, the 
coefficients of this expansion will be integer values. Moreover, we will 
assume the field strengths to have dimensions of $(\textrm{length})^{-1}$ 
so the moduli fields will be dimensionless.

Plugging this into eq.(\ref{super}), one can easily check that the 
superpotential is a cubic polynomial of the moduli
\begin{align}
W & = e_0
+ ih_0 S + \sum_{i=1}^3 [(ie_i - a_i S - b_{ii}U_i -\sum_{j\not= i}
b_{ij}U_j)T_i - i h_iU_i] \nonumber \\[0.2cm] & - q_1 T_2 T_3 -q_2
T_1 T_3 -q_3 T_1 T_2 + i m T_1 T_2 T_3\label{s2}
\end{align}

In addition, the fluxes in general will contribute to the RR tadpoles 
\cite{vz,cfi}. The relevant piece of the ten dimensional action is given by
\begin{equation}
\int_{M_4 \times T^6}[C_7 \wedge (m \overline{H}_3 +
d\overline{F}_2 )] + \sum_a N_a \int_{M_4 \times \Pi_a} C_7 \label{tadpol1}
\end{equation}
where we are considering the possibility of having stacks of $N_a$
D6-branes wrapping factorized 3-cycles
\begin{equation}
\Pi_a=(n_a^1, m_a^1)\otimes(n_a^2, m_a^2) \otimes(n_a^3, m_a^3)
\end{equation}

Plugging our background into (\ref{tadpol1}) then gives rise to the following 
tadpole cancellation conditions
\begin{align}
\sum_a N_a n_a^1 n_a^2 n_a^3 + \frac{1}{2}(h_0 m + a_1 q_1 + a_2 q_2 + a_3 q_3) &=
16\nonumber \\
\sum_a N_a n_a^1 m_a^2 m_a^3 + \frac{1}{2} (m h_1 - q_1 b_{11} - q_2 b_{21} -
q_3 b_{31}) & = 0 \nonumber \\
\sum_a N_a m_a^1 n_a^2
m_a^3 + \frac{1}{2}( m h_2 - q_1 b_{12} - q_2 b_{22} - q_3 b_{32}) & = 0 \label{tadpolo}\\
\sum_a N_a m_a^1 m_a^2 n_a^3 + \frac{1}{2} (m h_3 - q_1 b_{13} - q_2 b_{23} -
q_3 b_{33}) & = 0 \nonumber
\end{align}

Note that since the metric fluxes modify the homology of the
original torus, in principle not every 3-cycle $\Pi_a$ will be
a consistent cycle on which to wrap the D6-branes. Considerations 
about consistency for D6-branes in presence of metric fluxes will 
be made in section 4.

The vacuum structure of the scalar potential induced by (\ref{kahler}) 
and (\ref{s2}) has been analyzed in detail in \cite{cfi}. There, one 
can see how the fluxes induce a broad landscape of vacua consisting 
of both Minkowski and AdS vacua with broken or unbroken supersymmetry. 
Here we will briefly summarize some of the main results.

\subsection{Supersymmetric Minkowski models}

For the particular case of supersymmetric Minkowski vacua it was
found in \cite{cfi} that it is required the presence of metric fluxes and $m=0$.
In addition, generically one finds a large number of flat
directions, being able to fix at most three linear combinations of
complex moduli. The fluxes contribute to the RR tadpoles with the same
sign as D6-branes do, or do not contribute. This last possibility
represents a novelty with respect to the Gukov-Vafa-Witten
scenarios, where the fluxes always contributed positively to the RR tadpoles.

\subsection{No-scale models}

As 'no-scale' we distinguish models in which the superpotential is
independent of three of the moduli, so the explicit form
of the K\"ahler potential (\ref{kahler}) guarantees a cancellation
of the cosmological constant at tree-level, even though
supersymmetry is broken \cite{noscale}. These minima has very similar
properties to the ones found for the Minkowski models. In particular, again there is
a large number of moduli which remain unstabilized and the fluxes
always contribute to the RR tadpoles as D6-branes.

\subsection{AdS models}

This kind of vacua posses a different qualitative behavior than
the Minkowski ones. In particular, generically all
the moduli are stabilized but some linear combinations of complex
structure axions. As it will be shown along next section, this is actually 
required for consistency with the presence of D6-branes. Moreover, in
the cases with non vanishing metric fluxes, one has examples on
which the fluxes do not contribute to the RR tadpoles or
contribute with the same sign as the orientifold planes do. This
provides us with new possibilities for model building. In
particular, in section 5 we will see a concrete example on which
the tadpole contributions of the setup of branes are cancelled by
the effect of the fluxes, without the aid of an extra orbifold
twist.

\section{D6-branes in presence of fluxes}

Since the inclusion of metric fluxes modifies the homology, there will be 3-cycles of the original
torus which are no longer closed in the presence of metric fluxes. In this section we will see how 
there can appear inconsistencies in the low energy effective action when wrapping
D6-branes on these submanifolds.

In fact, the gauge fields living in the worldvolume of a stack of D6-branes
couple to the closed string axions through the following piece of
the action
\begin{equation}
\int_{M_4\times \Pi_a}(C_3\wedge F_a\wedge F_a+C_5\wedge F_a)=
\sum_{I=0}^3\int_{M_4}[p_I^a(\textrm{Im}~U_I)F_a\wedge 
F_a+N_ac_I^aC_I^{(2)}\wedge F_a]
\end{equation}
where $C_I^{(2)}$ is the Poincar\'e dual in 4d to $\textrm{Im}~U_I$ and 
the coefficients $\{p_I^a,c_I^a\}$ are given in terms of the wrapping 
numbers as
\begin{align*}
c_0^a&=m_a^1m_a^2m_a^3 & c_1^a&=m_a^1n_a^2n_a^3 & c_2^a&=n_a^1m_a^2n_a^3 & c_3^a&=n_a^1n_a^2m_a^3 \\
p_0^a&=n_a^1n_a^2n_a^3 & p_1^a&=n_a^1m_a^2m_a^3 & p_2^a&=m_a^1n_a^2m_a^3 & p_3^a&=m_a^1m_a^2n_a^3
\end{align*}

These couplings have revealed to play an important role in the 
cancellation of mixed $U(1)$ gauge anomalies through diagrams on 
which the gauge bosons exchange complex structure (and dilaton) 
axions \cite{alda}. In this sense, the scalars 
$\phi^a=\sum_Ic_I^a\textrm{Im}~U_I$ behave as Goldstone bosons, 
giving St\"uckelberg masses to the corresponding $U(1)$'s and 
transforming with a shift under the gauge transformations. We 
have seen however that the NSNS and metric fluxes induce terms 
in the superpotential which are linear in Im $U_I$ and therefore 
generically do not respect these transformation properties. Thus, 
in order to restore consistency and gauge invariance one has to 
impose the following condition on the 3-cycle $\Pi_a$ which the 
brane is wrapping
\begin{equation}
\int_{\Pi_a}(\overline{H}_3+dJ_c)=0 \label{consisten}
\end{equation}
In this way, the linear combinations of axions which become massive 
due to the Green-Schwarz mechanism will be orthogonal to the ones 
which become massive due to the effect of the fluxes. This constraint 
actually constitutes the generalization of the Freed-Witten anomaly 
cancellation condition to the case of a twisted torus and in fact, 
when $\omega=0$, one recovers the usual constraint for an ordinary torus.

Given a background for the NSNS 3-form and metric fluxes, the condition 
(\ref{consisten}) will determine what are the 3-cycles on which it is 
consistent to wrap the stacks of D6-branes. In the particular case of AdS 
vacua, this imposes $\Pi_a$ to be a special lagrangian cycle, so for this 
kind of models the closed string background restricts the brane content to 
supersymmetric configurations of D6-branes (i.e. with all Fayet-Iliopoulos 
terms vanishing) \cite{cfi}.

\section{An example of MSSM-like vacua}

As an illustration of all the above ideas, we will reproduce here a 
particular example of $N=1$ MSSM-like vacua with all the closed string 
moduli stabilized in AdS, which was already presented in \cite{cfi}.

We will consider a general isotropic background of NSNS, RR and metric 
fluxes given by $b_{ji}=-b_{ii}=b_i, a_i=a, q_i=-c_2$ and $e_i=c_1$ so 
there is just one overall K\"ahler moduli $T_1=T_2=T_3=T$ and the 
superpotential reduces to
\begin{equation}
W=e_0+3ic_1T+3c_2T^2+imT^3+ih_0S-3aST-\sum_{k=0}^3(ih_k+b_kT)U_k
\end{equation}
This choice of metric fluxes automatically satisfies (\ref{constra}).

We are interested in the supersymmetric AdS minima of the induced scalar 
potential. These can be obtained by imposing the vanishing of all 
the F-terms, i.e. $D_TW=D_{U_i}W=0$, but $W\neq 0$. With this one gets 
that all the real parts of the moduli are fixed in terms of the fluxes
\begin{equation}
a\textrm{Re}~S=2\textrm{Re}~T(c_2-m\textrm{Im}~T) \quad ; \quad 
3a\textrm{Re}~S=b_k\textrm{Re}~U_k \quad ; \quad 3a^2(\textrm{Re}~T)^2=
5h_0^2\lambda(\lambda+\lambda_0-1)
\end{equation}
where $\lambda$ is determined by the following cubic equation
\begin{equation}
160\lambda^3+186(\lambda_0-1)\lambda^2+27(\lambda_0-1)^2\lambda+
\lambda_0^2(\lambda_0-3)+\frac{27a^2}{mh_0^3}(e_0a-c_1h_0)=0
\end{equation}
and $\lambda_0=3c_2a/(mh_0)$.

Concerning the imaginary parts, the reader can check that the K\"ahler 
axions are completely fixed by the fluxes whereas only a linear 
combination of complex structure and dilaton axions is stabilized
\begin{equation}
\textrm{Im}~T=(\lambda+\lambda_0)\frac{h_0}{3a}\quad ; \quad 
3a\textrm{Im}~S+\sum_{k=1}^3b_k\textrm{Im}~U_k=3c_1+
\frac{3c_2}{a}(3h_0-7a\textrm{Im}~T)-\frac{3m}{a}\textrm{Im}~T(3h_0-8a\textrm{Im}~T)
\end{equation}

Additionally, the following constraint is required in order to stabilize 
the real parts of the moduli at positive values
\begin{equation}
3h_ka+h_0b_k=0 \quad \quad k=1,2,3 \label{otra}
\end{equation}

Let us concentrate now in the particular case of $c_2=2-h_i$, $a_i=16$, 
$m=b_i=4$ and $h_0=-12h_i$, which automatically satisfies eq.(\ref{otra}). 
The reader can check that choosing the fluxes in such a way the 
tadpole conditions read
\begin{equation}
\sum_aN_an_a^1n_a^2n_a^3=64\quad ; \quad \sum_aN_an_a^im_a^jm_a^k=-4 
\quad i\neq j\neq k \label{tadpo}
\end{equation}

Moreover, for large values of $h_0$ one has that the four dimensional 
dilaton scales as $e^{\phi_4}\simeq h_0^{-2}$, the ten dimensional one 
as $e^{\phi}\simeq h_0^{-1/2}$ and the flux density as $h_0^{-1/2}$. 
Thus, taking $h_0>>1$ one may stabilize the moduli at regions where 
the $\alpha'$ and $g_s$ corrections are under control and the effective 
supergravity approach remains valid.\footnote{Quantum corrections may 
play however an important role in the lifting of these vacua to dS, as
discussed in \cite{uno,dos}.}

One can add now the following supersymmetric setup of D6-branes 
\cite{cfi,cremades} in order to satisfy (\ref{tadpo})

\begin{table}[h]
\begin{center}
\begin{tabular}{|c||c|c|c|}
\hline $N_i$ & $(n_i^1,m_i^1)$ & $(n_i^2,m_i^2)$ & $(n_i^3,m_i^3)$ \\
 \hline\hline $N_a=4$ & $(1,0)$ & $(3,1)$ & $(3 , -1)$ \\ $N_b=1$ &
 $(0,1)$ & $ (1,0)$ & $(0,-1)$ \\ $N_c=1$ & $(0,1)$ & $(0,-1)$ &
 $(1,0)$ \\ \hline \hline $N_{h_1}=3$ & $(2,1)$ & $(1,0)$ & $(2,-1)$
 \\ $N_{h_2}=3$ & $(2,1)$ & $ (2,-1)$ & $(1,0)$ \\ $N_o=4$ & $(1,0)$ &
 $(1,0)$ & $(1,0)$ \\ \hline \end{tabular}
\end{center} 
\caption{A MSSM-like model with tadpoles cancelled by fluxes. Branes 
$h_1$, $h_2$ and $o$ are added in order to cancel the RR tadpoles.}
\end{table}

After separating some of the branes and after two of the $U(1)$'s get 
St\"uckelberg masses, the gauge group becomes 
$SU(3)\times SU(2)_L\times U(1)_R\times U(1)_{B-L}\times[U(1)\times SU(3)^2]$ 
with three generations of quarks and leptons, two doublets of Higgsses 
and extra matter fields involving the auxiliary branes $h_1, h_2$ and 
$o$. More details about the spectrum can be found in \cite{cfi}. In this way, 
we have constructed an example of consistent $N=1$ AdS vacuum with all 
the closed string moduli fixed and semi-realistic chiral spectrum.

\section{Discussion}

We have reviewed along the lines of \cite{cfi} some of the main properties 
for the vacua of simple $T^6/(\Omega(-1)^{F_L}\sigma)$ Type IIA 
orientifolds with non trivial RR, NSNS and metric fluxes. Unlike the Type 
IIB case, the richness of the flux options leads to a full stabilization 
of all closed string moduli in AdS without the need of non-perturbative 
effects. Moreover, one can find vacua on which the flux contribute to the 
RR tadpoles with the same sign as the O6-planes do or even do not contribute, 
so there is more freedom to find consistent models with all the moduli 
stabilized at the perturbative regions of the scalar potential. 

An interesting property of these models is that geometric fluxes survive 
the orientifold projection. These modify the homology of the original torus 
and constrains the open string sector. From the point of view of the low 
energy effective theory, this is reflected in the appearance of possible 
gauge inconsistencies in the worldvolume of the D6-branes. In the particular 
case of AdS models, this enforces the setup of D6-branes to be supersymmetric. 
As an illustration of these ideas, we have presented an example of $N=1$ 
MSSM-like vacua with all the closed string moduli stabilized in AdS.

However, still a lot of work has to be done in order to fully understand 
Type IIA compactifications with non trivial background fluxes. This involves 
the description of the ten dimensional supergravity solutions outside the 
approach of effective supergravity, the understanding of the geometric 
deformations beyond the toroidal geometry or the lift of these vacua to dS. 
We hope to come back to all these issues in the near future.

\begin{acknowledgement}
I would like to thank very especially the other two authors of \cite{cfi}, 
Anamaria Font and Luis E. Ib\'a\~nez, for their support and collaboration. As 
well I would like to thank F. Marchesano for useful discussions and the organizers of the Corfu 
2005 Summer Institute for the invitation to present this work. This 
contribution is supported by the Ministerio de Educaci\'on y Ciencia 
(Spain) through a FPU grant and by the European Commission under the 
RTN European Program MRTN-CT-2004-503369.
\end{acknowledgement}


\begin{thebibliography}{10}
\bibitem{grana} See e.g. M.~Grana,
  arXiv:hep-th/0509003 for a recent review and references.

\bibitem{gkp}  S.~B.~Giddings, S.~Kachru and J.~Polchinski,
  Phys.\ Rev.\ D {\bf 66} (2002) 106006
  [arXiv:hep-th/0105097].

\bibitem{cfi}  P.~G.~Camara, A.~Font and L.~E.~Ibanez,
  JHEP {\bf 0509}, 013 (2005)
  [arXiv:hep-th/0506066].

\bibitem{kashani} S.~Kachru and A.~K.~Kashani-Poor,
  JHEP {\bf 0503}, 066 (2005)
  [arXiv:hep-th/0411279].

\bibitem{louis}  T.~W.~Grimm and J.~Louis,
  Nucl.\ Phys.\ B {\bf 718}, 153 (2005)
  [arXiv:hep-th/0412277].

\bibitem{vz}  G.~Villadoro and F.~Zwirner,
  JHEP {\bf 0506}, 047 (2005)
  [arXiv:hep-th/0503169].

\bibitem{dwolf}  O.~DeWolfe, A.~Giryavets, S.~Kachru and W.~Taylor,
  JHEP {\bf 0507}, 066 (2005)
  [arXiv:hep-th/0505160].

\bibitem{cvetic} K.~Behrndt and M.~Cvetic,
  Phys.\ Rev.\ Lett.\  {\bf 95}, 021601 (2005)
  [arXiv:hep-th/0403049]; Nucl.\ Phys.\ B {\bf 708}, 45 (2005)
  [arXiv:hep-th/0407263].

\bibitem{kounnas} J.~P.~Derendinger, C.~Kounnas, P.~M.~Petropoulos and F.~Zwirner,
  Nucl.\ Phys.\ B {\bf 715}, 211 (2005)
  [arXiv:hep-th/0411276]; Fortsch.\ Phys.\  {\bf 53}, 926 (2005)
  [arXiv:hep-th/0503229].

\bibitem{kalop}  N.~Kaloper and R.~C.~Myers,
  JHEP {\bf 9905}, 010 (1999)
  [arXiv:hep-th/9901045].

\bibitem{ferrara}  G.~Dall'Agata and S.~Ferrara,
  Nucl.\ Phys.\ B {\bf 717}, 223 (2005)
  [arXiv:hep-th/0502066].

\bibitem{andri}  L.~Andrianopoli, M.~A.~Lledo and M.~Trigiante,
  JHEP {\bf 0505}, 051 (2005)
  [arXiv:hep-th/0502083].

\bibitem{hull}  C.~M.~Hull and R.~A.~Reid-Edwards,
  arXiv:hep-th/0503114.

\bibitem{scherk}  J.~Scherk and J.~H.~Schwarz,
  Phys.\ Lett.\ B {\bf 82} (1979) 60;  Nucl.\ Phys.\ B {\bf 153} (1979) 61.

\bibitem{pope} I.~V.~Lavrinenko, H.~Lu and C.~N.~Pope,
  Class.\ Quant.\ Grav.\  {\bf 15} (1998) 2239
  [arXiv:hep-th/9710243]. 

\bibitem{micu}  S.~Gurrieri, J.~Louis, A.~Micu and D.~Waldram,
  Nucl.\ Phys.\ B {\bf 654} (2003) 61
  [arXiv:hep-th/0211102].

\bibitem{new}  S.~Kachru, M.~B.~Schulz, P.~K.~Tripathy and S.~P.~Trivedi,
  JHEP {\bf 0303} (2003) 061
  [arXiv:hep-th/0211182].

\bibitem{noscale}  E.~Cremmer, S.~Ferrara, C.~Kounnas and D.~V.~Nanopoulos,
  Phys.\ Lett.\ B {\bf 133} (1983) 61.

\bibitem{alda}  L.~E.~Ibanez, R.~Rabadan and A.~M.~Uranga,
  Nucl.\ Phys.\ B {\bf 542} (1999) 112
  [arXiv:hep-th/9808139];  G.~Aldazabal, S.~Franco, L.~E.~Ibanez, R.~Rabadan and A.~M.~Uranga,
  J.\ Math.\ Phys.\  {\bf 42} (2001) 3103
  [arXiv:hep-th/0011073].

\bibitem{cremades}  D.~Cremades, L.~E.~Ibanez and F.~Marchesano,
  JHEP {\bf 0307} (2003) 038
  [arXiv:hep-th/0302105];  D.~Cremades, L.~E.~Ibanez and F.~Marchesano,
  JHEP {\bf 0405} (2004) 079
  [arXiv:hep-th/0404229].

\bibitem{uno}  V.~Balasubramanian and P.~Berglund,
  JHEP {\bf 0411} (2004) 085
  [arXiv:hep-th/0408054].

\bibitem{dos}   F.~Saueressig, U.~Theis and S.~Vandoren,
  Phys.\ Lett.\ B {\bf 633}, 125 (2006)
  [arXiv:hep-th/0506181].

\end{thebibliography}
\end{document}